\documentclass[amsmath,amssymb,superscriptaddress,nobalancelastpage,prl,twocolumn]{revtex4-1}

\usepackage{graphicx}
\usepackage{varioref}
\usepackage{xr-hyper}
\usepackage{xcolor}
\usepackage{hyperref}
\hypersetup{colorlinks,linkcolor=blue,urlcolor=blue,citecolor=blue}
\usepackage{ulem}
\usepackage{braket}

\newcommand{\RuSi}{URu$_2$Si$_2$}

\begin{document}

\title{Pressure-Induced Rotational Symmetry Breaking in URu$_2$Si$_2$}

\author{J. Choi}
\affiliation{Physik-Institut, Universit\"{a}t Z\"{u}rich, Winterthurerstrasse 190, CH-8057 Z\"{u}rich, Switzerland}

\author{O. Ivashko}
\affiliation{Physik-Institut, Universit\"{a}t Z\"{u}rich, Winterthurerstrasse 190, CH-8057 Z\"{u}rich, Switzerland}

\author{N. Dennler}
\affiliation{Physik-Institut, Universit\"{a}t Z\"{u}rich, Winterthurerstrasse 190, CH-8057 Z\"{u}rich, Switzerland}

\author{D. Aoki}
\affiliation{Universit\'e Grenoble Alpes, CEA, INAC-PHELIQS, 38000 Grenoble, France}
\affiliation{Institute for Materials Research, Tohoku University, Oarai, Ibaraki 311-1313, Japan}

\author{K. von Arx}
\affiliation{Physik-Institut, Universit\"{a}t Z\"{u}rich, Winterthurerstrasse 190, CH-8057 Z\"{u}rich, Switzerland}

\author{S. Gerber}
\affiliation{Laboratory for Micro and Nanotechnology, Paul Scherrer Institut, CH-5232 Villigen PSI, Switzerland}

\author{O. Gutowski}
\affiliation{Deutsches Elektronen-Synchrotron DESY, 22603 Hamburg, Germany.}


\author{M.~H.~Fischer}
\affiliation{Physik-Institut, Universit\"{a}t Z\"{u}rich, Winterthurerstrasse 190, CH-8057 Z\"{u}rich, Switzerland}
\affiliation{Institute for Theoretical Physics, ETH Z\"{u}rich, 8093 Z\"{u}rich, Switzerland }

 \author{J.~Strempfer}
\affiliation{Deutsches Elektronen-Synchrotron DESY, 22603 Hamburg, Germany.}

 \author{M.~v.~Zimmermann}
\affiliation{Deutsches Elektronen-Synchrotron DESY, 22603 Hamburg, Germany.}

\author{J. Chang}
\affiliation{Physik-Institut, Universit\"{a}t Z\"{u}rich, Winterthurerstrasse 190, CH-8057 Z\"{u}rich, Switzerland}

\begin{abstract}
Phase transitions and symmetry are intimately linked. Melting of ice, for example, restores translation invariance. The mysterious hidden order (HO) phase of \RuSi\ has, despite relentless research efforts, kept its symmetry breaking element intangible. Here we present a high-resolution x-ray diffraction study of the \RuSi\ crystal structure as a function of hydrostatic pressure. Below a critical pressure threshold $p_c\approx3$~kbar, no tetragonal lattice symmetry breaking is observed even below the HO transition $T_{HO}=17.5$~K. For $p>p_c$, however, a pressure-induced rotational symmetry breaking is identified with an onset temperatures $T_{OR}\sim 100$~K. The emergence of an orthorhombic phase is found and discussed in terms of an electronic nematic order that appears unrelated to the HO, but with possible relevance for the pressure-induced antiferromagnetic (AF) phase. Existing theories describe the HO and AF phases through an adiabatic continuity of a complex order parameter. Since none of these theories predicts a pressure-induced nematic order,  our  finding adds an additional symmetry breaking element to this long-standing problem.
\end{abstract}


\maketitle
  
Magnetism, superconductivity and the hidden order (HO) phase in \RuSi\ have been the subject of intense research~\cite{MydoshRMP2011,IkedaNatPhys2012,HauleNatPhys2009,GallagherNatComm2016,WiebeNatPhys2007,Santander-SyroNatPhys2009,ChandraNAT2013,RessouchePRL2012,RiggsSCI2015}. In particular, the symmetry breaking element associated with the hidden order lacks unequivocal evidence~\cite{WangPRB2017,Kung15Science,TonegawaNatComm2014,OkazakiSCI2011}. One influential set of theories describes the hidden order phase and magnetism through an adiabatic continuity of a single complex order parameter~\cite{IkedaNatPhys2012,HauleNatPhys2009,ChandraNAT2013}. Experimental explorations of the hydrostatic pressure and magnetic field phase diagrams are therefore paramount to solve this conundrum. Hydrostatic and chemical pressure tuning has established how the hidden order can be switched into a long-range antiferromagnetic (LRAF) phase~\cite{HassingerPRB2008,DasPRB2015,ButchPRB2010}. In fact, a modest pressure 
(reducing the lattice parameter by a few per mille) is sufficient to switch between the HO and LRAF ground states. Similarly, application of a high magnetic field ($\sim$35 T) along the $c$ axis quenches the HO into a spin-density-wave (SDW) phase~\cite{CorreaPRL2012,KnafoNatComm2016,JoPRL2007}. The putative adiabatic continuity between hidden order and magnetism implies that the entire pressure and magnetic field phase diagrams should be scrutinized. In fact, even though hydrostatic pressure compresses the unit cell volume~\cite{NiklowitzPRL2010}, the effect on the crystal lattice symmetry has not been elucidated. As the lattice and electronic degrees of freedom are coupled, it is of great interest to determine the crystal structure~\cite{KambePRB2018} across the \RuSi\ phase diagram.

\begin{figure}
\center{\includegraphics[width=0.5\textwidth]{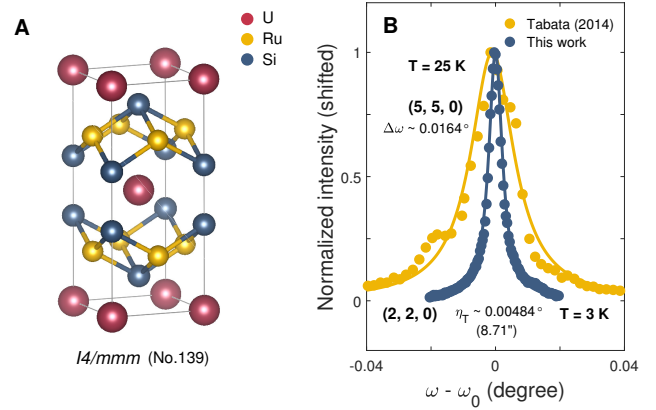}} 
\caption{Crystal structure and mosaicity of URu$_2$Si$_2$. (A) Ambient-pressure conventional unit cell of URu$_2$Si$_2$ with tetragonal \textit{I4/mmm} structure. (B) Transverse diffraction scans (sample rotation $\omega$) through $(h,h,0)$ Bragg peaks with $h$  being an integer as indicated. Blue (yellow) symbols indicate data from this work (Ref.~\cite{TabataPhilMag2014}). The Lorentzian peak width ${\eta}_{T}$ results from a combination of crystal mosaicity and instrument resolution.}\label{fig1}
\end{figure}

\begin{figure*}
\center{\includegraphics[width=0.9\textwidth]{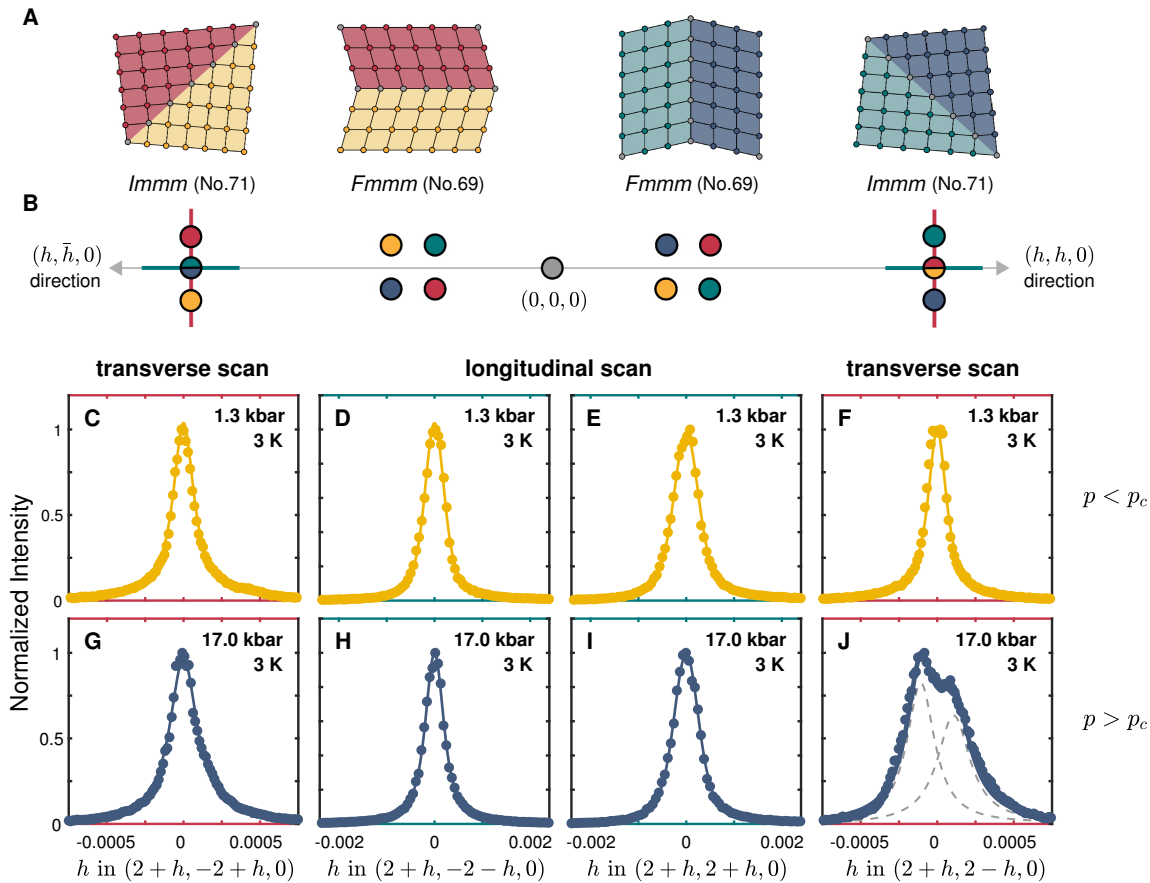}} 
\caption{(Color Online) Pressure-induced orthorhombicity in URu$_2$Si$_2$. 
(A) Possible domain formation for  \textit{Immm} and \textit{Fmmm} orthorhombic structures. (B) Corresponding Bragg peak 
along the orthorhombic $(h,\bar{h},0)$ and $(h,h,0)$ directions projected to a horizontal axis. The different domains lead to specific Bragg peak splittings as indicated by the color code. (C-J) Transverse and longitudinal scans through $(2,\bar{2},0)$ and $(2,2,0)$ for $p<p_c$ and $p>p_c$ with $p_c=3$~kbar. 
Transverse and longitudinal scans are fitted with Lorentzian and Voigt profiles,  respectively (solid lines). The transverse splitting of the $(2,2,0)$ reflection is modeled by fitting two Lorentzians (solid and dashed lines).}\label{fig2}
\end{figure*}

\begin{figure*}[t]
\center{\includegraphics[width=0.85\textwidth]{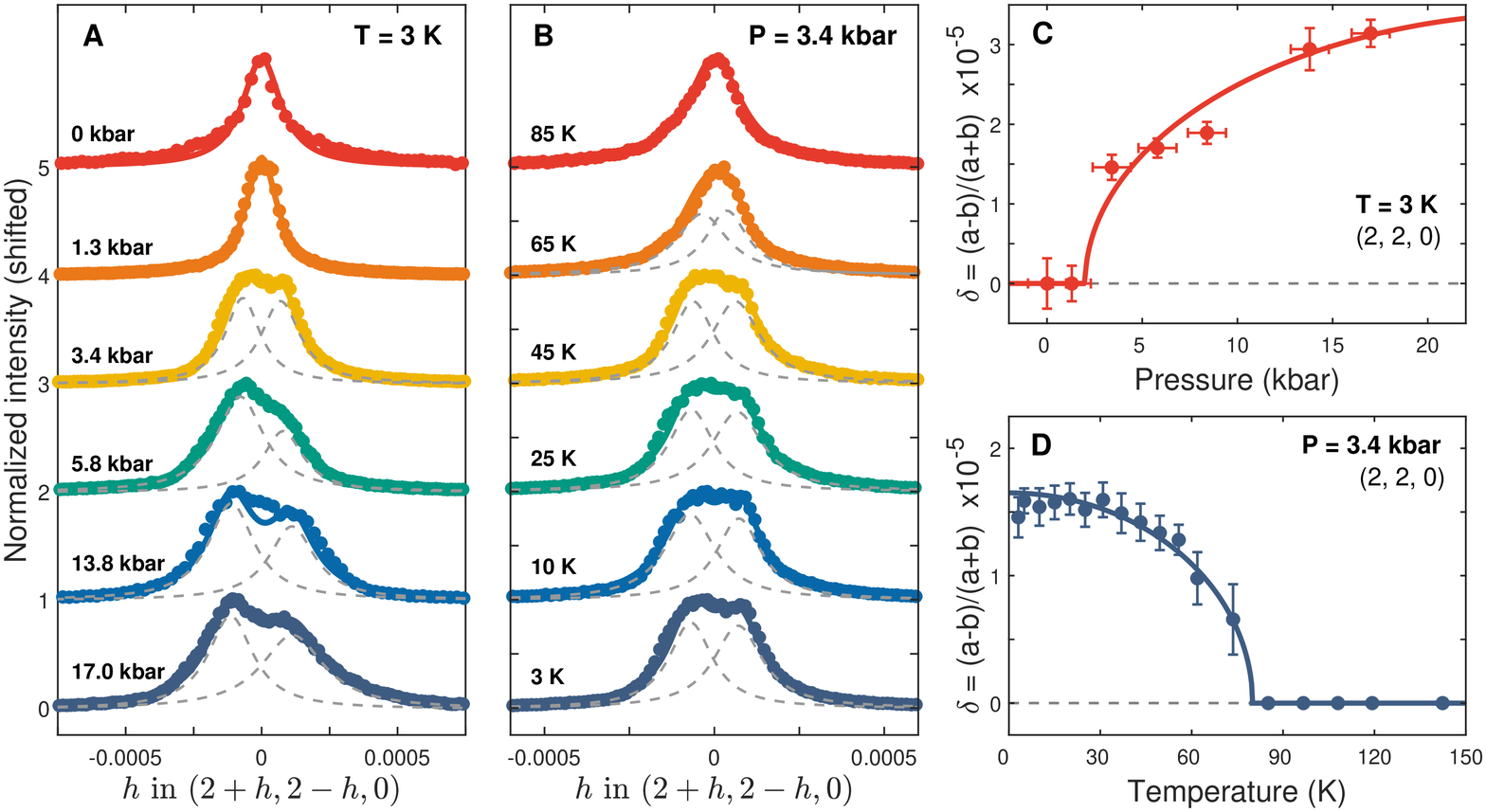}} 	
\caption{Temperature and pressure dependence of Orthorhombicity. (A,B) Transverse scans through the $(2,2,0)$ Bragg reflection for hydrostatic pressures and temperatures as indicated. Dashed grey lines indicate the results of fitting with two Lorentzian peaks and solid lines are their sum. (C) Hydrostatic-pressure and (D) temperature-dependent orthorhombic order parameter $\delta$, derived from the Lorentzian fits shown in (A) and (B). Solid lines are guides to the eye. Error bars on $\delta$ are set by the standard deviation of the relevant fitting parameters. Systematic errors of the pressure are indicated by horizontal bars.}	
\label{fig3}
\end{figure*} 

Here, we present a hard x-ray diffraction study of the URu$_2$Si$_2$ 
crystal structure as a function of hydrostatic pressure. A single crystal with pristine mosaicity was selected. At ambient pressure, the crystal structure remains tetragonal across the hidden order transition and down to the lowest measured temperatures (3 K). Above a critical pressure, $p_c=3$~kbar, an orthorhombic phase is identified. The orthorhombic onset temperature $T_{OR}\sim 100$ K is, after an initial dramatic rise, only weakly pressure dependent. It is discussed whether the associated electronic nematic order parameter is a trigger (or consequence) of the orthorhombic transition. The weakness of the orthorhombic order parameter in comparison to the onset temperature suggests that the rotational symmetry breaking is electronically driven and that the lattice follows as a secondary effect. From the topology of the established phase diagram, the hidden and nematic orders appear 
uncorrelated. Nematicity may, however, be a precondition for magnetism. In fact, none of the adiabatic continuity models predicts a pressure-induced nematic order. As such, our findings provide a different symmetry breaking element to the problem.

A high-quality single crystal ($\sim1\times1\times1$~mm$^3$) was selected for hard x-ray diffraction experiments under hydrostatic pressure. This URu$_2$Si$_2$ crystal is from a batch that has previously been used for scattering \cite{RessouchePRL2012,WalkerPRB2011} and quantum oscillation \cite{HassingerPRL2010,AokiJPSJ2012} experiments. The residual-resistivity-ratio (RRR) value of these crystals is typically in the range 100-500 \cite{HassingerPRL2010,AokiJPSJ2012}. Our studies were carried out at the P07 triple-axis diffractometer at PETRA III (DESY-Hamburg) using 100-keV x rays in transmission scattering geometry. A 18-kbar piston pressure cell \cite{Zimmermann08RCI,Huecker10PRL,IvashkoSCIREP2017} with standard Daphne oil as the pressure medium and a La$_{1.875}$Ba$_{0.125}$CuO$_4$~\cite{Huecker10PRL} crystal for pressure calibration (see Supplementral Fig. 1 \cite{Suppl}) was used. The pressure cell was cooled by a helium cryostat with a crystal orientation allowing access to the $(h,k,0)$ scattering plane. In this fashion, the $c$ and piston axes are parallel and hence there is no geometric inequivalence between the $a$- and $b$-axis directions. Weak in-plane uniaxial pressure can therefore be excluded entirely. Scattering vectors are specified in tetragonal reciprocal notation with ambient-pressure (3-K) lattice parameters $a=b=4.123$ and $c=9.58$~\AA. We checked that the temperature dependence of the in-plane lattice parameter is consistent with previous neutron scattering experiments~\cite{NiklowitzPRL2010} (see Supplemental Fig. 2 \cite{Suppl}).

To investigate crystal structure [Fig.~\ref{fig1}(A)], high-quality single crystallinity (quantified by mosaicity) and excellent instrumental resolution are required. Figure \ref{fig1}(B) displays a transverse scan through the $(2,2,0)$ Bragg reflection of our URu$_2$Si$_2$ crystal. A Voigt fit reveals a negligible Gaussian contribution and a Lorentzian half width at half maximum (HWHM) $\eta_T=9"$ (1.5$\times$10$^{-4}$~\AA$^{-1}$)  defining the resolution along that direction.  This resolution is finer than  previous studies~\cite{TabataPhilMag2014,TonegawaNatComm2014} (Fig.~\ref{fig1}). Along the longitudinal direction through $(2,2,0)$, our setup has comparable Gaussian $\sigma_L=3.7\times$10$^{-4}$~\AA$^{-1}$ and $\eta_L=3.8\times$10$^{-4}$~\AA$^{-1}$ contributions. The high-temperature crystal structure of \RuSi\ belongs to the \textit{I4/mmm} space group~\cite{NandiPRL2010,TonegawaNatComm2014}. This tetragonal structure has 15 nonisomorphic subgroups for which two (\textit{Fmmm} and \textit{Immm}) are orthorhombic~\cite{TonegawaNatComm2014}. The possible domains of these two orthorhombic structures are shown in Fig.~\ref{fig2}(A). 
Corresponding Bragg peak splittings are schematically illustrated in Fig.~\ref{fig2}(B) along the $(h,h,0)$ and $(h,\bar{h},0)$ reciprocal directions. The relative Bragg peak intensities depend on the exact domain population. Provided sufficient experimental resolution, longitudinal ($2\theta$) and transverse ($\omega$) scans through $(h,h,0)$ and $(h,\bar{h},0)$ Bragg peaks are adequate to distinguish between the \textit{Fmmm} and \textit{Immm} structures, as shown in Fig.~\ref{fig2}. The \textit{Fmmm} structure splits the Bragg peak in both the transverse and longitudinal direction whereas only a transverse splitting is expected for \textit{Immm}.

\begin{figure}[t]
\center{\includegraphics[width=0.5\textwidth]{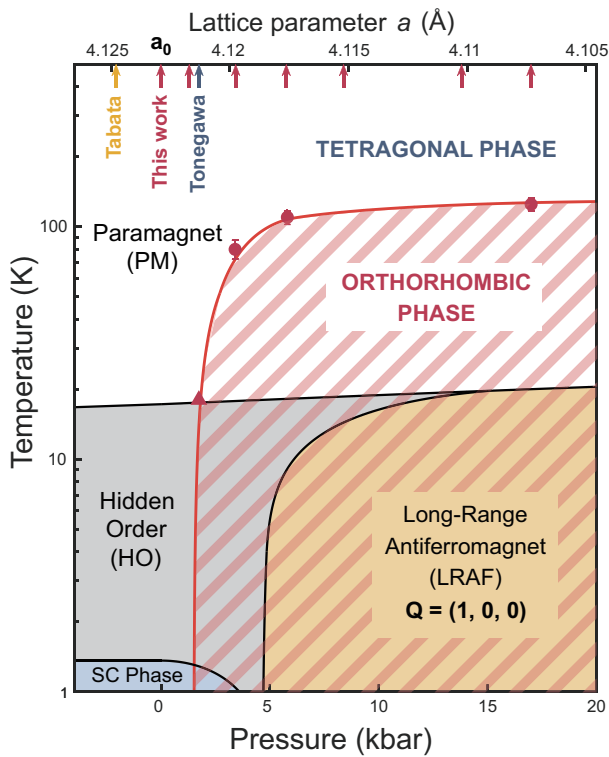}} 	
\caption{Temperature-pressure phase diagram of URu$_2$Si$_2$. The hidden order, long-range antiferromagnet order, and superconducting (SC) phases are displayed with gray, yellow and blue shadings, respectively. In addition, the tetragonal-to-orthorhombic phase boundary is indicated by the hashed area. The emergence of orthorhombicity is given both as a function of pressure and in-plane lattice parameter. By contrast, HO, LRAF and SC are indicated only as a function of pressure (Ref. \cite{HassingerPRB2008}). Notice that the in-plane lattice parameter $a$ is an absolute scale whereas the pressure axis remains relative without an ambient-pressure lattice parameter $a_0$ determination. The orthorhombic transition (circular markers) was only measured for a subset of pressures applied. The triangular marker indicates the orthorhombic \textit{Fmmm} onset found by Tonegawa \textit{et al.} (Ref.~\onlinecite{TonegawaNatComm2014}). Due to the logarithmic temperature scale, the absence of orthorhombicity at pressures of 0 and 1.3 kbar is not displayed.}	\label{fig4}
\end{figure}

The absence of longitudinal and transverse $(2,2,0)$ and $(2,\bar{2},0)$ Bragg peak splittings for $p<p_c=3$~kbar suggests that the system remains tetragonal even inside the hidden order phase [Fig.~\ref{fig2}(C)-\ref{fig2}(F)].
By contrast, for $p>p_c$ a transverse splitting of the $(2,2,0)$ Bragg peak is observed [Fig.~\ref{fig2}(J)]. The fact that $(2,\bar{2},0)$ remains sharp indicates a highly polarised domain population. At $p=17$~kbar, the transverse splitting amounts to $\approx1.3\sigma_L$. Our resolution is therefore good enough to resolve a longitudinal \textit{Fmmm} splitting (if it existed). As the $p=1.3$ and 17~kbar longitudinal Bragg peaks are essentially identical [Figs.~\ref{fig2}(E),~\ref{fig2}(I) and Supplemental Fig. 3 \cite{Supp}], the high-pressure orthorhombic structure is of \textit{Immm}-type. The onset of orthorhombicity is revealed by transverse scans through $(2,2,0)$ versus temperature and pressure [Figs. \ref{fig3}(A) and \ref{fig3}(B)]. The orthorhombic order parameter is defined as $\delta=(a-b)/(a+b)$ where $a$ and $b$ are in-plane lattice parameters extracted by fitting the  Bragg peak splitting \cite{McIntyrePRB1988}.
A double Lorentzian fit with the widths set by the resolution $\eta_T$ was used. The orthorhombic order parameter $\delta$ as a function of pressure and temperature is shown in Figs.~\ref{fig3}(C) and \ref{fig3}(D), respectively. The pressure-dependent onset temperature of orthorhombicity, defined by $\delta>0$, is compared to the phase space of the hidden order and antiferromagnetic state in Fig. \ref{fig4}.  

Next, we comment on the fact that ambient-pressure orthorhombicity has previously been reported in ultrapure ($RRR=600$) URu$_2$Si$_2$~\cite{TonegawaNatComm2014}. To this end, it is useful to consider the in-plane lattice parameter in detail. The pressure phase diagram ($0\rightarrow 20$~kbar) of URu$_2$Si$_2$ corresponds to a 20 per mille tuning of the in-plane lattice parameter. Literature-quoted low-temperature ambient-pressure in-plane lattice parameters $a_0$ vary by 5 per mille \cite{PalstraPRL1985,TonegawaNatComm2014,TabataPhilMag2014} -- almost 20\% of the pressure phase diagram. If the ambient-pressure in-plane lattice constant is not precisely determined, this translates into a large error bar in the pressure phase diagram. Quoting exact lattice parameters is therefore important when discussing orthorhombicity and magnetism. The ambient-pressure orthorhombicity reported by Tonegawa \textit{et al.}~\cite{TonegawaNatComm2014} is, for example, found in a crystal with a lattice parameter corresponding to a finite hydrostatic pressure within our reference frame. As such, there is no discrepancy between the reports in that regard. Two central differences are, however, that Tonegawa \textit{et al.}~\cite{TonegawaNatComm2014} reported
(i) a \textit{Fmmm} orthorhombic structure that (ii) coincides with the HO onset temperature. Our high-pressure diffraction results are consistent with an \textit{Immm} orthorhombic structure. though, near the pressure onset of orthorhombicity $p_c$ our resolution is not sufficient to distinguish between \textit{Immm} and \textit{Fmmm}. We can therefore not exclude an additional \textit{Fmmm} phase near $p_c$. For the orthorhombic onset temperature no correlation with the hidden order phase is found even near $p_c$. We notice that ultrasound experiments in low magnetic fields report a dominant softening of the $C_{11}-C_{12}$ mode \cite{YanagisawaPRB2013,YanagisawaPRB2018} -- consistent with a transition to the \textit{Immm} space group. Furthermore, the temperature onset of the $C_{11}-C_{12}$ softening at 120 K is consistent with the appearance of the \textit{Immm} structure in our diffraction experiment ($T_{OR}\sim100$~K). Although an additional \textit{Fmmm} structure may occur near the low-temperature tetragonal-to-orthorhombic transition, we conclude that the \textit{Immm} structure is dominating the pressure phase diagram (Fig. \ref{fig4}). 

An interesting question is whether the orthorhombicity is elastic or electronic driven. It is worth noticing that in contrast to \RuSi, many quasi-two-dimensional systems are pushed toward higher symmetry upon application of hydrostatic pressure. For example, it is typically the case for transition metal oxides with a high-temperature \textit{I4/mmm} structure \cite{Huecker10PRL}. This trend is also found in isostructural SrFe$_2$As$_2$ \cite{WuSciRep2014} and dichalcogenides such as TaS$_2$ \cite{SiposNMAT2008}. The fact that symmetry in \RuSi\ is lowered with hydrostatic pressure suggests the underlying physics is different. Another remarkable difference is that the orthorhombic order parameter $\delta$ of \RuSi\ is at least an order of magnitude smaller than what is found, for example, in pnictide systems \cite{NandiPRL2010,KothapalliNatComm2016}. Yet, the onset temperatures are comparable. This is suggestive of an electronic nematic ordering parameter being the primary and the lattice orthorhombicity a secondary consequence. Notice that to detect this nematic order parameter directly, for example, with resistivity requires single domain crystals with a sufficiently short in-plane lattice parameter.
 
Finally, the softening of the $C_{11}-C_{12}$ ultrasound mode, consistent with an  \textit{Immm} structure, has already been discussed in terms of hybridization between the uranium $5f$ orbitals and the conduction electrons \cite{YanagisawaPRB2013,YanagisawaPRB2018}. Stronger hybridization favors a more pronounced softening. Hydrostatic pressure reduces the unit cell volume that in turn enhances all hybridizations including those of uranium $5f$ and conduction electrons.
This provides an electronic ("Band-Jahn-Teller") mechanism \cite{YanagisawaPRB2013,YanagisawaPRB2018} for the $C_4\rightarrow C_2$ lattice symmetry breaking.


The topology of the phase diagram (Fig.~\ref{fig4}) suggests no obvious connection between the nematic and hidden order parameters. Since both nematicity and long-range antiferromagnetic (LRAF)~\cite{AmitsukaJMM2007} order are pressure induced, a coupling between two is not inconceivable. We note that the pressure onset of LRAF has not been experimentally calibrated to the in-plane lattice parameter scale. It is therefore not impossible that nematicity and LRAF have identical onset pressure. The high-pressure onset temperature of LRAF order seems to coincide with that of the HO parameter. This has led to a class of theories describing the HO and LRAF within a single complex order parameter connected through an adiabatic continuity \cite{ChandraNAT2013,IkedaNatPhys2012,HauleNatPhys2009,RessouchePRL2012}. 
In fact, a plethora of order parameters has been suggested, where some multipolar orders can break $C_4$ down to $C_2$ on the lattice level \cite{IkedaNatPhys2012,ThalmeierPRB2011}, some break $C_4$ to $C_2$ but only in the spin channel \cite{ChandraNAT2013}, and then there is the suggestion of an arrested Kondo effect \cite{HauleNatPhys2009}, or chiral density wave \cite{Kung15Science} that does not break rotational symmetry. However, for all these cases, rotation symmetry is at best broken in the HO phase, but never in the LRAF phase. Our experimental findings are therefore adding an entirely different electronic symmetry breaking element to the problem. Future work will clarify whether nematicity is part of a complex order or whether it is triggering the adiabatic switching between antiferromagnetism and the hidden order.
The authors are grateful to P.W.J. Moll, S. Benhabib, M. Janoschek, R. Flint, and H. Nojiri for inspiring discussions. J.C., O.I., and J.C. thank the Swiss National Science Foundation for support through the Grant No, BSSGI0$\_$155873. D. A. acknowledges support from the MEXT of Japan Grant-in-Aid for Scientific Research (through Grants No. JP15H05882, No. JP15H05884, No. JP15K21732, No. JP15H05745, and No. JP16H04006) and the European Research Council (through ERC-starting grant - NewHeavyFermion).





\bibliography{main_PRL}

\newcommand{\beginsupplement}{
        \setcounter{table}{0}
        \renewcommand{\thetable}{S\arabic{table}}
        \setcounter{figure}{0}
        \renewcommand{\thefigure}{S\arabic{figure}}}

\beginsupplement
\clearpage
\onecolumngrid

\section{Supplemental Material}

\begin{figure}[htb]
 	\begin{center}
 		\includegraphics[width=0.9\textwidth]{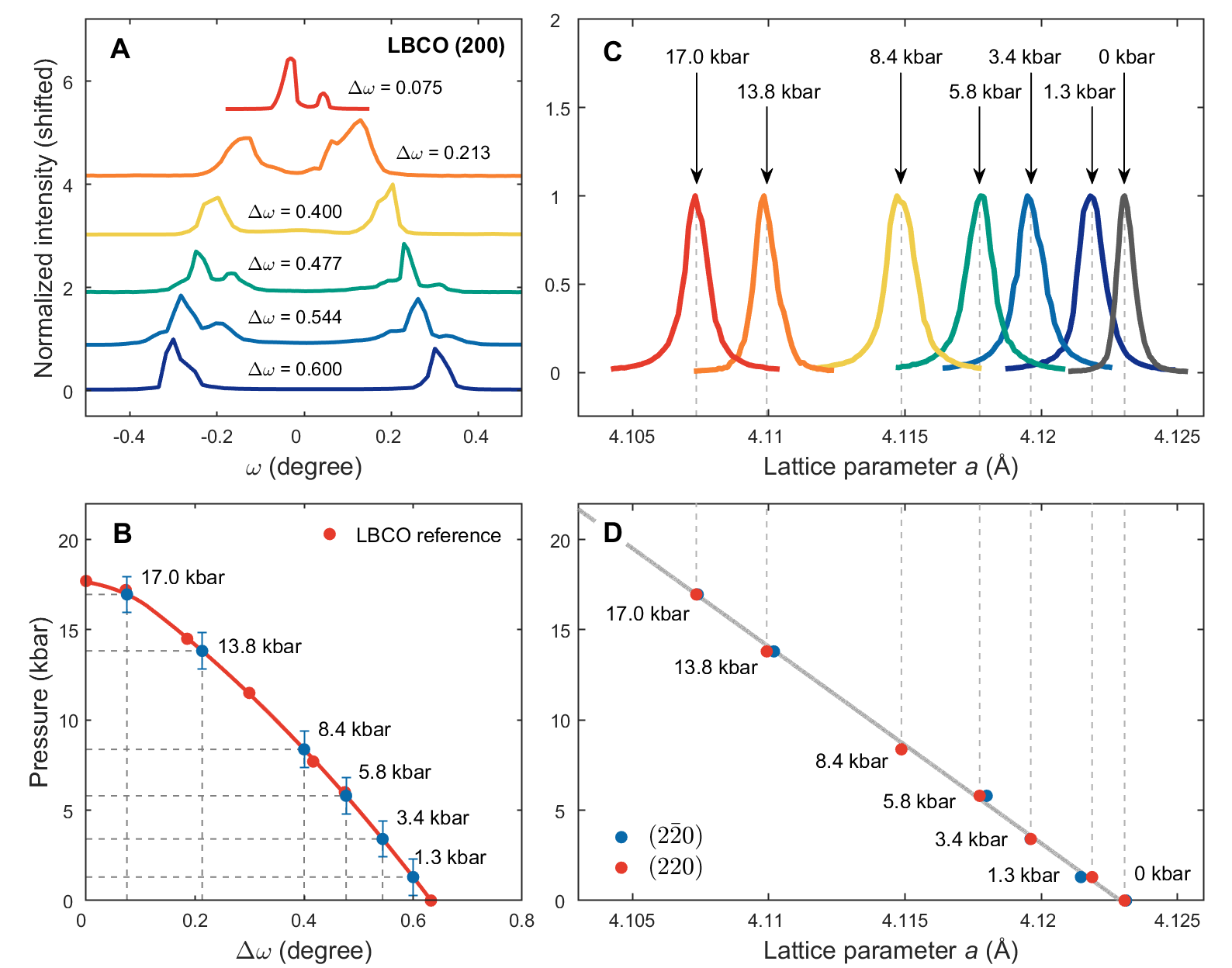}
 	\end{center}
 	\caption{Calibration of hydrostatic pressure and lattice constants. \textbf{(A)} Measurement of splitting $\Delta\omega$ due to the orthorhombic stuctural transition in transverse ($\omega$) scans of an La$_{1.875}$Ba$_{0.125}$CuO$_4$ (LBCO) crystal. \textbf{(B)} For calibration of hydrostatic sample pressure, the splittings $\Delta\omega$ are compared to the data from \cite{Huecker10PRL}. \textbf{(C)} Diffracted intensity measured along the longitudinal (2$\theta$) direction at 3 Kelvin for hydrostatic pressure as indicated. 
 	The horizontal axis has been converted into in-plane lattice parameter via Bragg law.
 \textbf{(D)} Linear relationship between  pressure and in-plane lattice parameter used for the horizontal axis in Fig.~\ref{fig4}.}\label{fig:S1}
\end{figure}

\begin{figure}[htb]
 	\begin{center}
 		\includegraphics[width=0.7\textwidth]{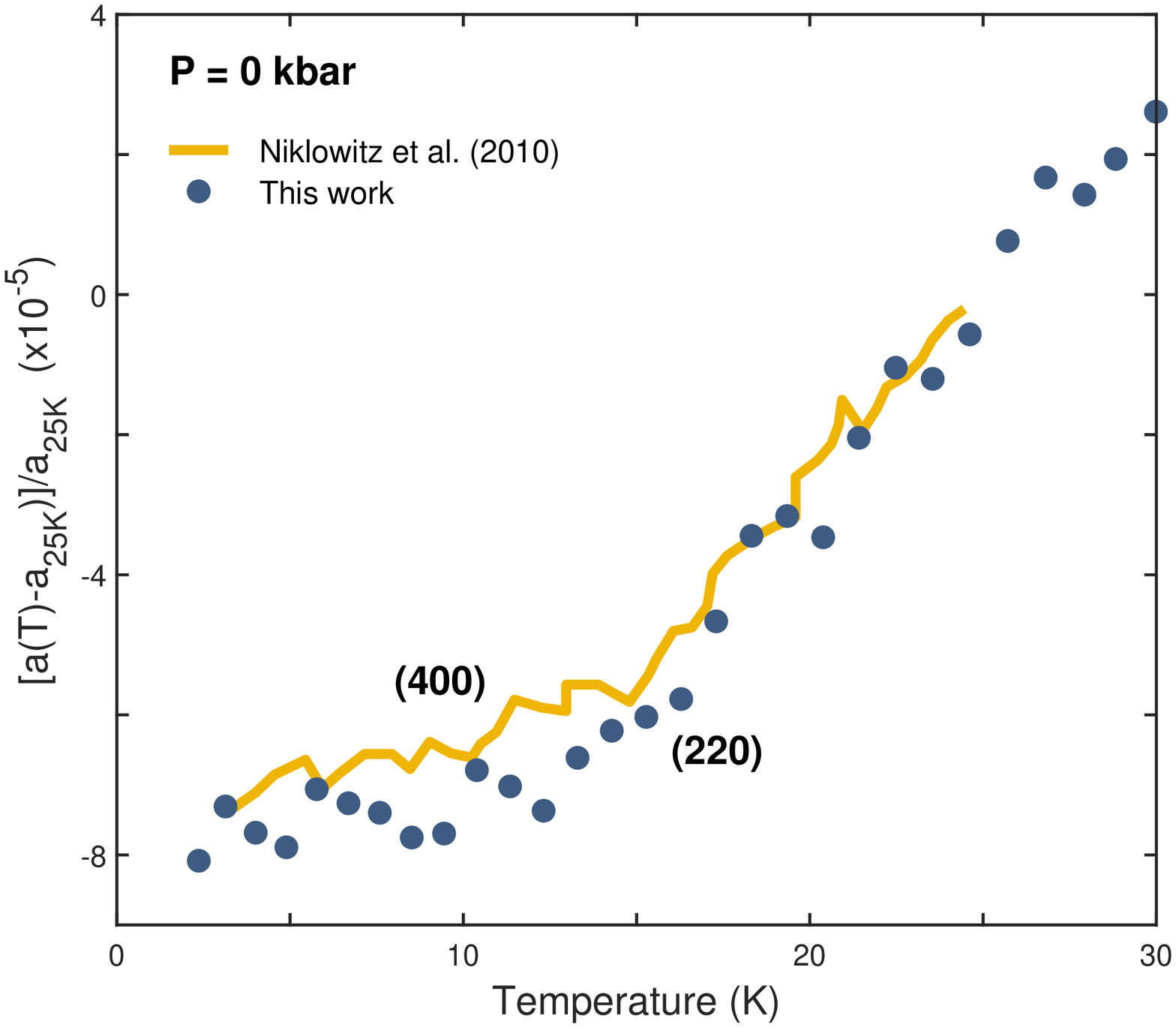}
 	\end{center}
 	\caption{Temperature dependence of the in-plane ambient pressure lattice parameter $a(T)$.
 	In-plane lattice parameter versus temperature. Solid point and yellow line are extracted from the (2,2,0) 
 	[this work] and (4,0,0) [Niklowitz \textit{et al}. (Ref.~\cite{NiklowitzPRL2010})] Bragg reflections. }
 	
\end{figure}

\begin{figure}[htb]
 	\begin{center}
 		\includegraphics[width=0.99\textwidth]{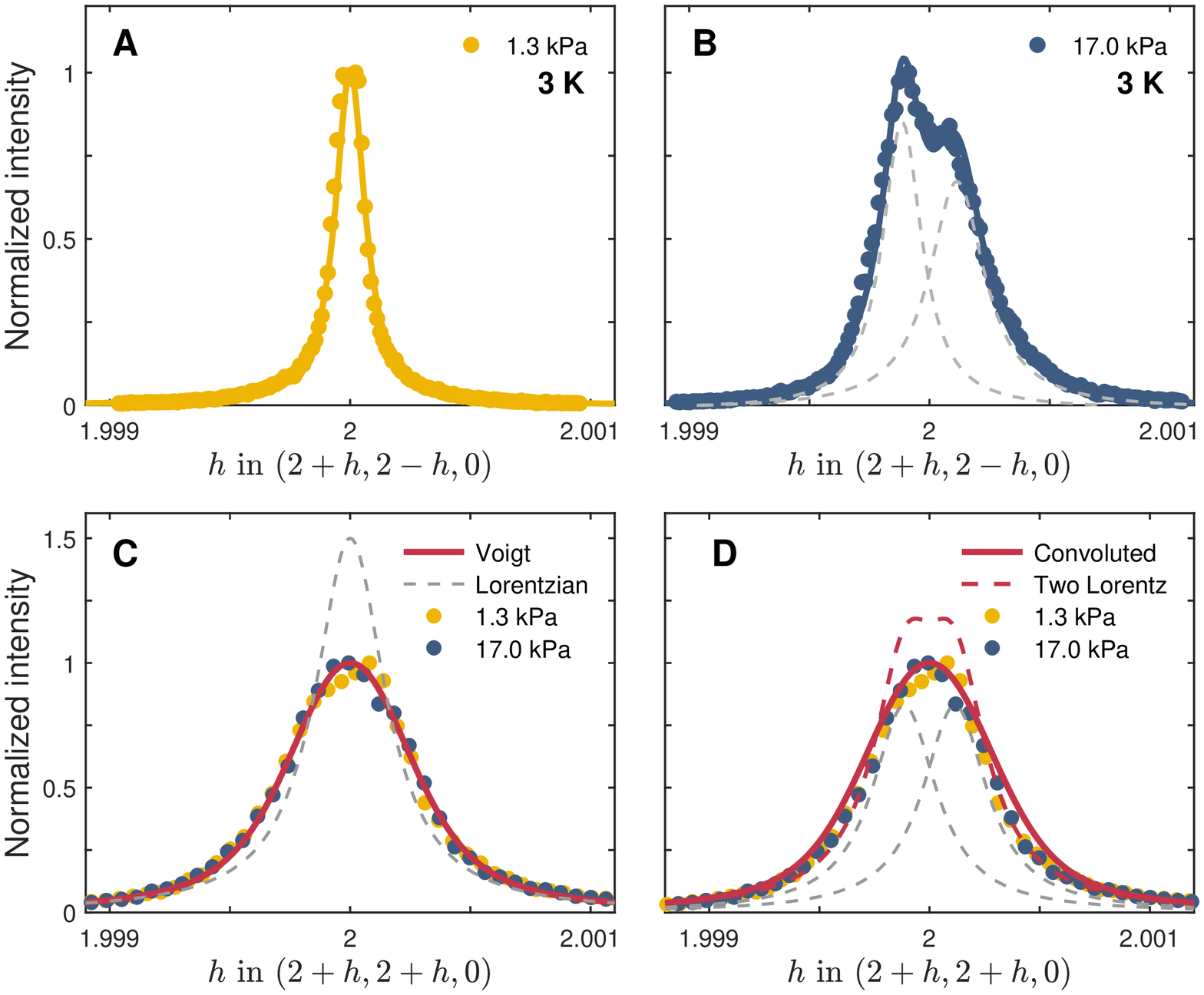}
 	\end{center}
 	\caption{Exclusion of the \textit{Fmmm} structure. (A,B) Transverse ($\omega$) scans through the (2,2,0) Bragg reflection for hydrostatic pressures and temperatures as indicated. Solid lines represent Lorentzian fits. In panel B, two Lorentzians are used as indicated by the dashed grey lines. (C,D) Longitudinal scans through (2,2,0). Data in C and D are identical and no discernible difference between $p=1.3$ and 17 kbar is observed. 
 	In panel C, the Bragg peak profile is modelled by a Voigt function with a single Lorentzian component shown by the grey dashed line. By contrast, in D a Gaussian convolution of a double Lorentzian is assumed. The splitting of the Lorentzians was set identical to that found in the transverse direction (panel B). The width of the two Lorentzians was set by the longitudinal Lorentzian component derived from the fit of the $p=1.3$~kbar. The modelling of C and D corresponds to the \textit{Immm} and \textit{Fmmm} orthorhombic structures respectively. As the \textit{Fmmm} modelling is not describing the observed profile, we conclude in favor of the \textit{Immm} space group.}
 	
\end{figure}

\end{document}